\documentclass[aps,prc,preprint,groupedaddress,showpacs]{revtex4}
\usepackage{graphicx}

\begin{document}

\title{Connection between the high energy-scale evolution of the P- and
T-odd $\pi N N$ coupling constant and the strong $\pi N N$ interaction}

\author{P.\ G.\ Blunden}
\affiliation{Department of Physics and Astronomy, University of Manitoba,
		Winnipeg, MB, Canada R3T 2N2}

\author{S.\ Kondratyuk}
\affiliation{Department of Physics and Astronomy, University of Manitoba,
		Winnipeg, MB, Canada R3T 2N2}

\begin{abstract}
The large energy-scale behaviour of the parity and time-reversal
violating (PTV) pion-nucleon coupling constant is analyzed in a model
combining renormalization-group techniques and the dressing of the PTV
vertex with a pion loop. With the strong $\pi N N$ vertex as a mixture
of the pseudovector and pseudoscalar couplings, we show that depending
on the admixture parameter, two qualitatively distinct types of
behaviour are obtained for the PTV coupling constant at high energy
scales: an asymptotic freedom or a fixed-point. We find a critical value
of the admixture parameter which delineates these two scenarios. Several
examples of the high-energy scale behaviour of the PTV $\pi N N$
constant are considered, corresponding to realistic hadronic models of
the strong pion-nucleon interaction.
\end{abstract}

\maketitle
\newpage

In addition to the usual strong interaction, pions and nucleons may have
a much suppressed Parity- and Time-reversal Violating (PTV) interaction
proportional to the Charge conjugation and Parity (CP) violating
$\theta$ term in the QCD Lagrangian~\cite{Khr97}. The upper bound for
the PTV $\pi N N$ coupling constant, denoted here as $c$, can be
extracted from measurements of Electric Dipole Moments (EDMs) of the
neutron and various atomic and molecular systems. It is found to be
extremely small: $c < 7 \times 10^{-12}$, corresponding to $|\theta| < 3
\times 10^{-10}$~\cite{Khr97,Pos05}. At higher energies, the PTV effects
may either remain suppressed, or alternatively they may grow leading to
a significant CP violation. Which of these two scenarios is more likely
to occur has important implications for possible extensions of the
Standard Model.

In a previous paper~\cite{Kon07} a hadronic model was developed in which
a PTV $\pi N N$ vertex was dressed with meson loops. Using a
renormalization-group approach, it was found that if one uses a purely
pseudoscalar strong $\pi N N$ vertex in the loop, the PTV coupling
vanishes at high energy scales (an asymptotic freedom). Alternatively,
by using a purely pseudovector strong $\pi N N$ vertex, one obtains a
PTV coupling constant tending to a fixed value at high energies. In this
report we generalize these results to the case where a more realistic
strong $\pi N N$ vertex is used in the loop integral for the PTV $\pi N
N$ coupling.

To obtain a good description of nucleon-nucleon and pion-nucleon
scattering, as well as pion photoproduction and other processes up to
resonance energies, one often uses a strong $\pi N N$ vertex as a
mixture of the pseudovector (PV) and pseudoscalar (PS) interactions:
\begin{equation}
\Gamma_S = g \left\{ \frac{a}{2m} \gamma_5 \slash\!\!\!{q} + (1-a) \gamma_5 \right\} \,,
\label{eq:strong_pinn}
\end{equation}
where $q$ is the pion four-momentum and $g \approx 13$ is the strong
pion-nucleon coupling constant. In different approaches, the PV-PS
mixing parameter $a$ appears either as a
constant~\cite{Gro92,Gro93,Gou94,Sur96} fitted to data, or as an
energy-dependent function which is determined from a phenomenological
fit, as in~\cite{Dre99}, or obtained from a self-consistent dressing of
the vertex, as in~\cite{Kon01}. 

\begin{figure}
\includegraphics[width=6in]{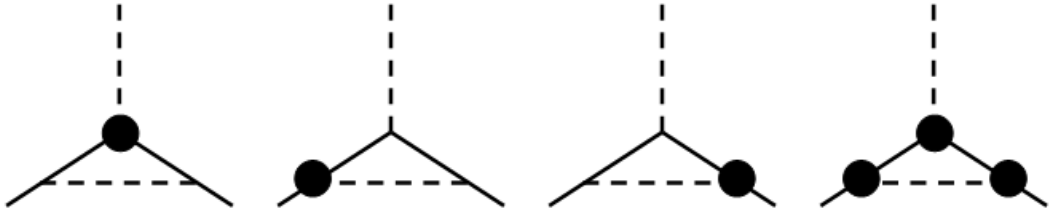}
\caption{Loop diagrams used to extract the beta-function. The solid and
dashed lines are nucleons and pions. The blob represents a PTV $\pi N N$
vertex described by the coupling constant $c$. The point vertex
represents the strong $\pi N N$ vertex given in
Eq.~(\ref{eq:strong_pinn}).
\label{fig:loops}}
\end{figure}

The energy-scale evolution of the PTV coupling constant can be studied
using renormalization-group techniques. The renormalization-group
equation for the dependence of the coupling constant $c$ on the energy
scale $\mu$ reads~\cite{Wei95}
\begin{equation}
\mu \frac{d c}{d \mu} = \beta[c] \,,
\label{eq:rn_eq}
\end{equation}
where $\beta[c]$ is the beta-function. In this report we restrict
ourselves to the system of nucleons and pions only while generalizing
the calculation of Ref.~\cite{Kon07} to a more general strong $\pi N N$
vertex given in Eq.~(\ref{eq:strong_pinn}), with an arbitrary parameter
$a$. Following the approach of~\cite{Kon07}, we extract the
beta-function from the one-loop diagrams depicted in
Fig.~(\ref{fig:loops}), within the standard procedures~\cite{Wei95} of
dimensional regularization and modified minimal subtraction. The
resulting beta-function can be written
\begin{equation}
\beta[c] = A c \, + B c^3 \,,
\label{eq:beta_fun}
\end{equation}
where
\begin{equation}
A = (5 a^2 - 2) \frac{g^2}{16 \pi^2} \, , \,\,\,\,\, B = - \frac{1}{8 \pi^2} \,.
\label{eq:ab}
\end{equation}
The renormalization-group equation (\ref{eq:rn_eq}) can be integrated
between energy scales $\mu_1$ and $\mu_2$, yielding the solution
\begin{equation}
\left( \frac{\mu_2}{\mu_1} \right)^A = \frac{c(\mu_2)}{c(\mu_1)} 
\sqrt{ \frac{1+{B \over A} c^2(\mu_1)}{1+{B \over A} c^2(\mu_2)} } .
\label{eq:solut}
\end{equation} 
Approximations to this solution for smaller and larger values of
$c(\mu)$ have simpler forms, as was discussed in some detail in
Ref.~\cite{Kon07}. 

It is well-known~\cite{Wei95} that, through the renormalization-group
equation (\ref{eq:rn_eq}), the sign of the beta-function determines the
behaviour of the PTV coupling constant $c(\mu)$ at large energy scales
$\mu$. Two qualitatively distinct types of behaviour were identified in
Ref.~\cite{Kon07}. In the first type $\beta[c] < 0$, $\forall c > 0$,
entailing $c(\mu \rightarrow \infty) \rightarrow 0$, i.~e.~an asymptotic
freedom behaviour. This scenario obtains if one chooses $a=0$ in
Eq.~(\ref{eq:strong_pinn}), thus using a purely pseudoscalar strong $\pi
N N$ vertex. If one uses a purely pseudovector vertex, setting $a=1$ in
Eq.~(\ref{eq:strong_pinn}), the second scenario is realized. In this
case the beta-function is positive for moderate values of $c$, in which
interval $\beta[c]$ increases, reaches a maximum value and then
decreases, becoming negative after crossing zero at a fixed point $c^*
\ne 0$. This behaviour leads to $c(\mu \rightarrow \infty) \rightarrow
c^*$, i.~e.~ the PTV coupling constant approaches the fixed point $c^*$
at sufficiently high energies. Since many realistic hadronic models of
strong pion-nucleon interaction at low and intermediate energies use a
mixture of the pseudovector and pseudoscalar vertices, a question arises
what kind of high-energy behaviour of the PTV coupling constant
corresponds to these hadronic models. We study this question in the
remainder of this report.

Assuming without loss of generality that $c>0$, the beta-function in
Eq.~(\ref{eq:beta_fun}) will be negative-definite if the PV-PS ratio $a
\le a_{crit}$, where the critical value is equal to
\begin{equation}
a_{crit}=\sqrt{2 \over 5} \approx 0.63 \,.
\label{eq:a_crit}
\end{equation}
This critical value allows us to establish a connection between hadronic
models with a strong $\pi N N$ interaction and the asymptotic behaviour
of the PTV $\pi N N$ coupling constant. Specifically, a model with $a
\le a_{crit}$ is consistent with the asymptotic freedom behaviour of the
PTV coupling constant, $c(\mu \rightarrow \infty) \rightarrow 0$,
whereas a model with $a > a_{crit}$ advocates the fixed point scenario,
$c(\mu \rightarrow \infty) \rightarrow c^*$. The fixed value $c^*$ is
determined by solving the equation $\beta[c^*]=0$. Using
Eq.~(\ref{eq:beta_fun}), we obtain 
\begin{equation}
c^*=g \sqrt{\frac{5 a^2 - 2}{2}} \,.
\label{eq:c_fixed}
\end{equation}
The two types of high-energy behaviour described above coincide when
$c^* \rightarrow 0$, which happens for $a \rightarrow a_{crit}$. 

The accompanying table is a compilation of results of several successful
hadronic models~\cite{Gro92,Gro93,Gou94,Sur96} in which the mixing
parameter $a$ in the strong $\pi N N$ vertex Eq.~(\ref{eq:strong_pinn})
is assumed to be constant. From the relation between $a$ and $a_{crit}$
of Eq.~(\ref{eq:a_crit}) we obtain the high-energy limit of the PTV
pion-nucleon coupling constant $c$, given in the last column of the
table (Eq.~(\ref{eq:c_fixed}), with $g \approx 13$, is used to determine
the values of $c^*$). 

\begin{center}
\begin{tabular}{|c|c|c|}
\hline
Model & $\approx a$ & $c(\mu \rightarrow \infty)$ \\
\hline
\cite{Gro92} Relativistic      &  $\; 0.78 \;$ & $c^* \approx 9.4 $  \\
\cite{Gro92} Non-relativistic  &  $\; 0.59 \;$ & $0$    \\
\cite{Gro93}                   &  $\; 0.75 \;$ & $c^* \approx 8.3 $  \\
\cite{Gou94}                   &  $\; 0.97 \;$ & $c^* \approx 15.1 $  \\
\cite{Sur96}                   &  $\; 0.8 \;$  & $c^* \approx 10.1 $  \\
\hline
\end{tabular}
\end{center}

The hadronic models in which the mixing parameter $a$ is a function of
energy are exemplified by Refs.~\cite{Dre99} and \cite{Kon01}. Although
the details of these two models are different, they both describe a
strong $\pi N N$ interaction which is essentially pseudovector at zero
energy, but with a pseudoscalar component increasing with energy. This
translates to a mixing parameter $a$ being unity at zero energy and
decreasing monotonously at higher energies. If at sufficiently high
energies $a$ becomes smaller than $a_{crit}$, then $c(\mu \rightarrow
\infty) \rightarrow 0$. Generally it is not known whether the models
\cite{Dre99} and \cite{Kon01} are applicable at energies sufficiently
high for the renormalization-group considerations. Therefore the
asymptotic behaviour $c(\mu \rightarrow \infty)$ corresponding to these
models cannot be determined unambiguously.

To summarize, in this report we have linked the PV-PS mixture in a
strong $\pi N N$ vertex with the type of the high-energy scale evolution
of the PTV $\pi N N$ coupling constant $c$. Using the one-loop
beta-function, we have found a critical values of the PV-PS admixture
parameter $a$, which separates the scenarios of asymptotic freedom and a
fixed point for $c(\mu \rightarrow \infty)$. For illustration, we have
calculated the fixed points for several hadronic models. At present, we
restricted our model to the nucleon and pion degrees of freedom only,
which made possible the direct connection with the hadronic models of
the strong $\pi N N$ interaction. This approach, however, is less
reliable at sufficiently high energies, where heavier mesons, as well as
quark-gluon degrees of freedom, are expected to become important.

\acknowledgments

This work was supported in part by NSERC (Canada).


\newpage

\end{document}